\documentclass[12pt, draftclsnofoot,onecolumn]{IEEEtran}
\usepackage{dsfont}
\usepackage{amssymb}
\usepackage{subfig}
\usepackage{graphicx, amssymb, amsmath}
\usepackage{multirow}
\emergencystretch=\maxdimen
\usepackage{tikz}
\newcommand*\ncircled[1]{\tikz[baseline=(char.base)]{
		\node[shape=circle,draw,inner sep=0.8pt] (char) {#1};}}

\IEEEoverridecommandlockouts
\begin{document}

\title{Vehicular Communication Networks in Automated Driving Era}

\author{Shan~Zhang,~\IEEEmembership{Member,~IEEE,}
        Jiayin~Chen,~\IEEEmembership{Student~Member,~IEEE,}
        Feng~Lyu,~\IEEEmembership{Student~Member,~IEEE,}
        Nan~Cheng,~\IEEEmembership{Member,~IEEE,}
        Weisen~Shi,~\IEEEmembership{Student~Member,~IEEE,}
        and~Xuemin~(Sherman)~Shen,~\IEEEmembership{Fellow,~IEEE}
\thanks{Shan~Zhang is with Beijing Key Laboratory of Computer Networks, the School of Computer Science and Engineering, Beihang University, Beijing, 100191, P.R.China. (Email: zhangshan18@buaa.edu.cn).}%
\thanks{Jiayin~Chen, Nan~Cheng, Weisen Shi and Xuemin~(Sherman)~Shen are with the Department of Electrical and Computer Engineering, University of Waterloo, 200 University Avenue West, Waterloo, Ontario, Canada, N2L 3G1 (Email:\{j648chen, n5cheng, w46shi, sshen\}@uwaterloo.ca).}
\thanks{Feng~Lyu is with the Department of Computer Science and Engineering, Shanghai Jiao Tong University, Shanghai, 200240, P.R.China. (Email: fenglv@sjtu.edu.cn).}
}

\maketitle

\begin{abstract}
	
	Embedded with advanced sensors, cameras and processors, the emerging automated driving vehicles are capable of sensing the environment and conducting automobile operation, paving the way to modern intelligent transportation systems (ITS) with high safety and efficiency.
	On the other hand, vehicular communication networks (VCNs) connect vehicles, infrastructures, clouds, and all other devices with communication modules, whereby vehicles can obtain local and global information to make intelligent operation decisions.
	Although the sensing-based automated driving technologies and VCNs have been investigated independently, their interactions and mutual benefits are still underdeveloped.
	In this article, we argue that VCNs have attractive potentials to enhance the on-board sensing-based automated vehicles from different perspectives, such as driving safety, transportation efficiency, as well as customer experiences.
	A case study is conducted to demonstrate that the traffic jam can be relieved at intersections with automated driving vehicles coordinated with each other through VCNs.
	Furthermore, we highlight the critical yet interesting issues for future research, based on the specific requirements posed by automated driving on VCNs.
	
\end{abstract}


\section{Introduction}

The idea of automated driving, which expects vehicles to drive without human participation, was proposed in 1956. 
However, it has been regarded as a fantasy for decades, and until recently the advancements in artificial intelligence (AI), electronic and information technologies are turning it into reality.
An automated driving vehicle should be capable of sensing surrounding environment with on-board cameras, sensors and transceivers, understanding the driving scenarios, and then making appropriate control decisions to operate engines, wheels and brakes.
With human beings liberated from driving tasks, automated driving technologies hold the promise to bring the revolutionary Intelligent Transportation Systems (ITSs), in terms of enhanced driving safety, transportation efficiency, and enriched travel experiences with a variety of on-road services.
Although the fully-automated vehicles are still underdeveloped, lower-level driver-assist automated systems have been commercialized, such as adaptive cruise control (ACC), collision warning and parking assistant.
By 2015, 15\% of the vehicles have been equipped with driver-assist automated systems, while around 50\%--60\% of the vehicles are expected to have higher-level automation by 2020, with estimated marketing value of 7 trillion US dollars.

Vehicular communication networks (VCNs), where vehicles are expected to communicate with any devices to enable interactions, cooperations and coordinations on road, are considered as another paradigm to the ITSs.
Although sharing the same goal, the existing studies on VCNs and on-board sensing-based automated driving are rather independent, without considering the potential of interdiscipline.
In fact, the sensing and communication technologies present complementary advantages, and the integration of VCNs can further enhance the performance and experiences of automated driving.
Specifically, the performance of automated driving systems can be greatly constrained by technical challenges of sensing range, machine version defects, training data sets, and limited processing capabilities.
VCNs can break these constraints by providing each individual vehicle global information of all aspects, unlimited computing and storage resources, as well as cooperation abilities.
In this way, automated vehicles are empowered with safety-enhanced individual driving intelligence, highly-efficient cooperative swarm intelligence, and customer-friendly serving intelligence.
As illustration, we conduct a case study on automated intersection passing, which demonstrates that the communication-based control can significantly improve transportation efficiency and reduce intersection passing delay, compared with the conventional traffic light based control. 

Despite of the attractive benefits, the sensor-rich and application-abundant automated driving vehicles also pose significant challenges to the design of VCNs, such as massive connectivity, high mobility, channel uncertainty, strict requirements on latency and reliability.
As the VCNs are still in infancy, how to address these issues needs extensive research efforts from all aspects.
The conventional problems, like network architecture design, management and performance analysis need to be revisited considering the specific features of automated driving.
In addition, new research topics may also arise in such highly-mobile cyber physical systems, such as the joint design of communication and control, security and privacy provisioning, outsourcing of computing and storage.

\section{Sensing-Based Automated Driving}
    \label{sec_architecture}
    
    \subsection{Automated Driving Technologies}
    
\begin{figure*}[!t]
    	\centering
    	\includegraphics[width=6in]{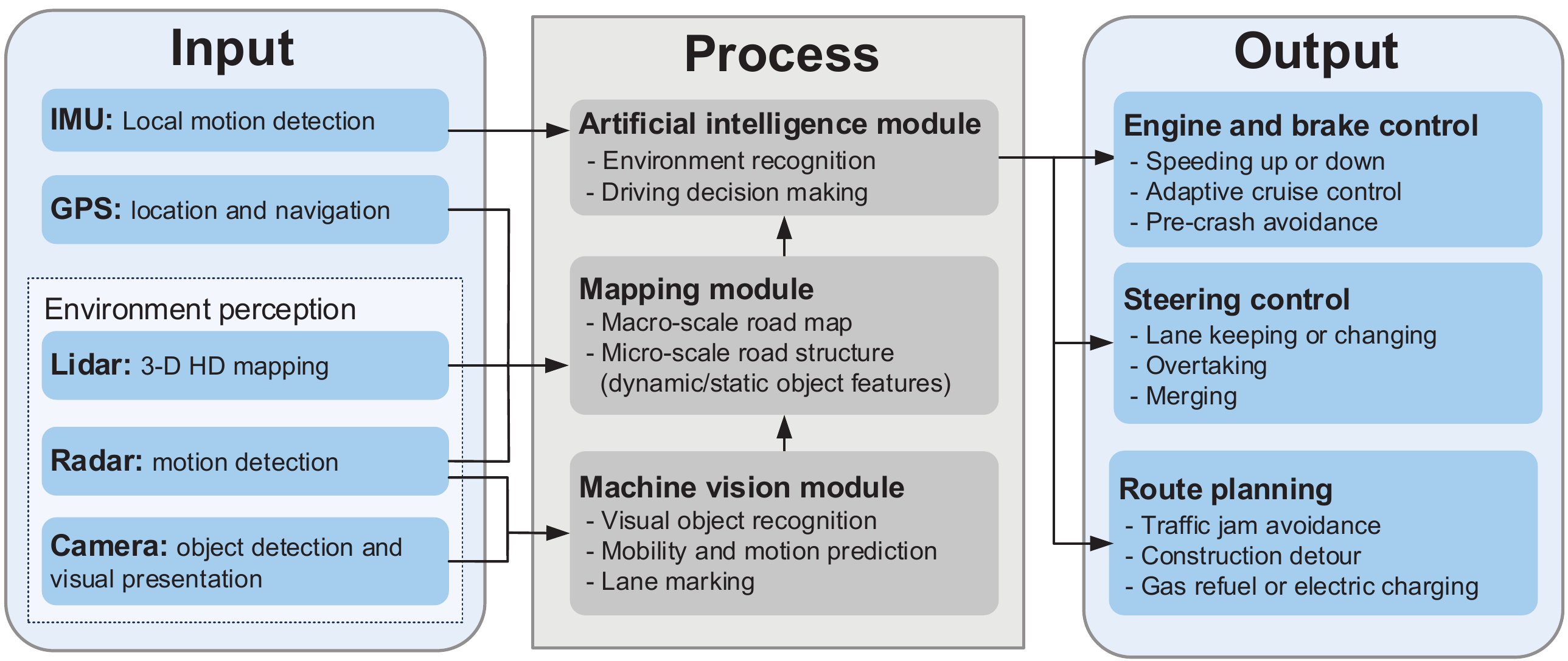}
    	\caption{Automated driving system and enabling technologies.}
    	\label{fig_automated_driving_system}
    \end{figure*}

    Fig.~\ref{fig_automated_driving_system} shows the automated driving systems.
    Three kinds of information are required for system input: (1) Global Positioning System (GPS) localization and navigation; (2) Inertial Measurement Unit (IMU); and (3) perception of surrounding driving environment.
    With the GPS receiver and IMU, vehicles get aware of their mobility information, such as position, speed, heading directions, acceleration, turn rate and inclination.	
    Sensing technologies, such as RAdio Detection and Ranging (RADAR), LIght Detection and Ranging (LIDAR), and optical cameras, enable vehicles with the perception of surrounding environment.
    RADAR sensors can be deployed at different parts of the vehicle to detect the motion measurement of all directions within range {{from}} 100 to 200 meters.
    LIDAR systems, using laser beams, are implemented to generate high-resolution 3-D road maps with spatial structures and detailed information, such as lanes, road features in addition to vehicles and obstacles.
    The optical cameras can present visible and colorful images which are most suitable for scene interpretation, object detection and classification.

    With complementary advantages, all the three kinds of input information are fused to analyze the driving environment and make the automobile operation decisions, through three functional components of machine vision, map construction, and AI-based decision making, as shown in Fig.~\ref{fig_automated_driving_system}.
    Machine vision can extract objects of interests by analyzing the raw camera images, such as traffic lights, lanes, pedestrians, road signs and facilities. 
    The processing result is fused with LIDAR and RADAR information, to construct structural 3-D map with both static and dynamic object properties of the surrounding environment.
    In addition to the micro-scale local 3-D road map, the macro-scale real-time map, containing the vehicle location, roads and corresponding traffic volumes, is also maintained by GPS signals for route planning and navigation.
    The AI-based decision making module acts as the human brain based on highly developed deep learning techniques, which jointly processes all gathered (and preprocessed) information to determine the action of the vehicle.
    Finally, the automobile control decisions will be transmitted to engines, brakes and wheels through wired signals, to realize automatic speed and steering control.  

    \subsection{Potential Obstacles}
    
    The landing of fully automated driving needs to overcome obstacles from all aspects, including technical challenges of on-board sensing and processing, privacy and security risks, marketing and regulation issues.
    
    On-board sensing technologies are constrained by detection {{range}} (RADRA with 100-200 meters, LIDAR with 60 meters), and may fail in bad weather conditions (camera in rainy or snowy days).
    In addition, the performance of automated driving highly depends on the data set.
    Accordingly, comprehensive training is required to develop a safe and efficient driving system, introducing high cost of road test.
    Still, the automated driving systems can make wrong decisions when the situation is out of data sets, like fading lanes or unexpected objects. 
    For instance, the automated driving system may crash and shut down when memory or processing capability is insufficient.
    
    Comparing with traditional vehicles, automated driving vehicles rely heavily on eletronic equipment, exposing users to risks of cyber-security attacks. 
    Once hacked, the vehicles can be controlled remotely, causing both individual and public safety concerns.
    As ultra-intelligent end devices, the automated vehicles can be utilized to collect data of any rider, such as the daily routes, voice, video, consumption records and entertainments during the trip. 
    These data, if obtained by hackers, may lead to privacy disclosure, like behavioral patterns, financial and even health conditions. 
    
    Although automated driving is promising to launch in taxi and logistics businesses to reduce the high cost of human resources, this may lead to job loss and political concerns.
    Furthermore, when human drivers are replaced by machine, transportation laws and policies should be amended to regulate the responsibility of car manufactures and owners for crashes, which is still under extensive disputes.

\section{Vehicular Communication Enhanced Intelligence}
    \label{sec_management}
    This section introduces how vehicular communications can enhance the automated driving intelligence, and reviews existing research works on VCN-assisted vehicle automation.
    
    \subsection{Vehicle Communication Technologies}
    
\begin{figure*}[!t]
    	\centering
    	\includegraphics[width=6in]{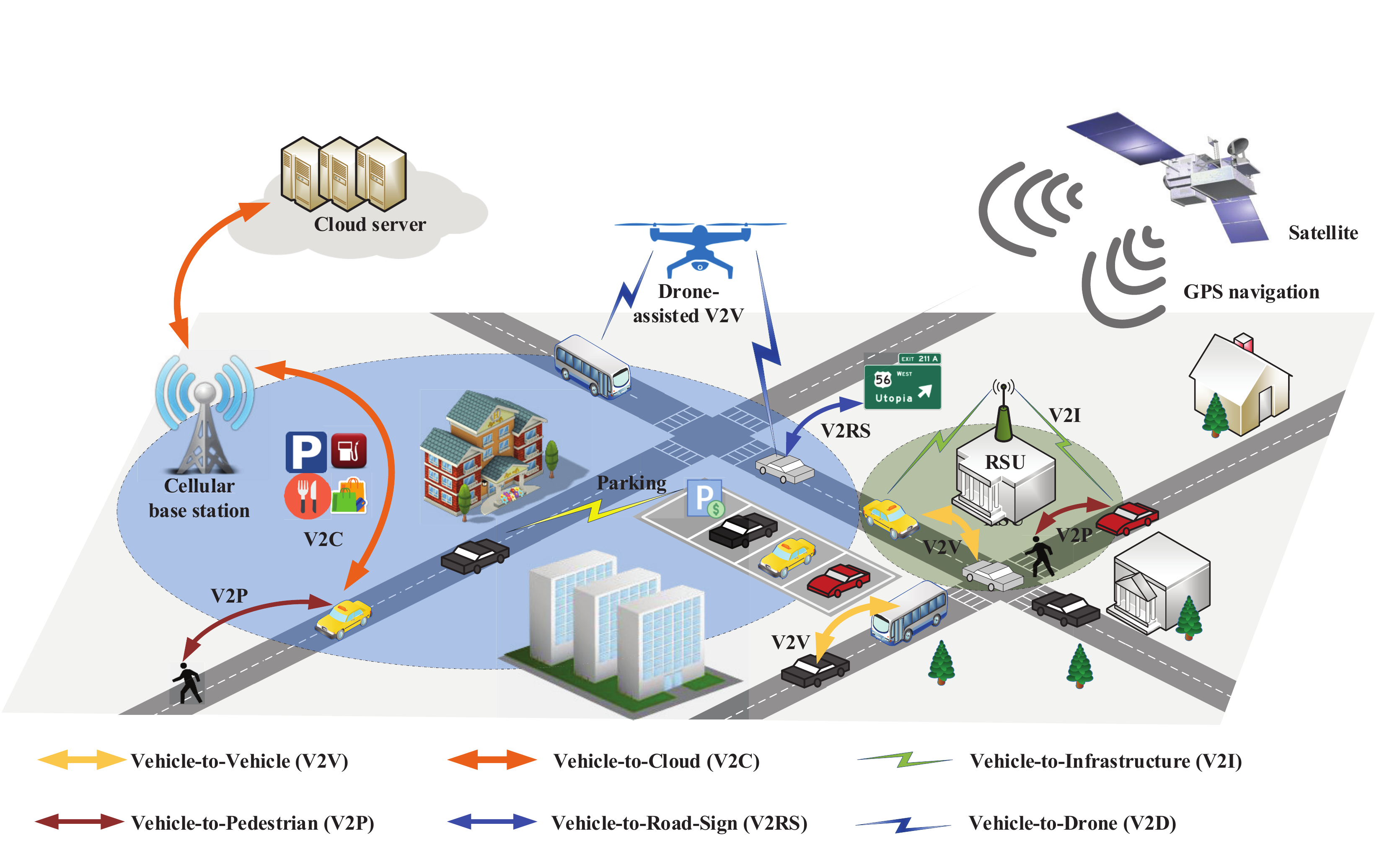}
    	\caption{{{VCN illustration with different V2X communication modes.}}}
    	\label{fig_V2X_scenario}
    \end{figure*}

    The initial VCNs mainly include two modes, vehicle-to-vehicle (V2V) and vehicle-to-infrastructure (V2I) communications.
    In specific, V2V communications allow vehicles to share information with each other (like emergency braking and accidents), while V2I communications can provide Internet access through road side units (RSUs).		
    With the emerging Internet of Things (IoT) technologies, vehicles are expected to communicate with any devices based on the vehicle-to-everything (V2X) communication technologies.
    A V2X communication network is illustrated in Fig.~\ref{fig_V2X_scenario}, including V2V, V2I, vehicle-to-pedestrians (V2P), vehicle-to-cloud (V2C), vehicle-to-road-signs (V2RS), and other devices with communication capabilities. 
    {{Vehicles need to select the appropriate communication modes to meet the quality of service (QoS) demand based on the network status.}}
    The connected vehicles market is expected to increase at a Compound Annual Growth Rate (CAGR) of 35.54 and achieve 37.7 million units by 2022, estimated to be 155.9 billion US dollars. 
    
    Considering the potential on-road mobile applications, VCNs are required to support diverse context information transmissions summarized in Table~\ref{tab_context_V2X}.
    Several potential solutions have been proposed to launch VCNs, including both WLAN-based IEEE 802.11p and cellular-based LTE-Vehicle (LTE-V).
    The 802.11p standard was initialized to realize dedicated short-range communications (DSRC) for V2V connections, and also developed the V2I mode to form a nationwide network through RSU access points.
    The Federal Communications Commission (FCC) of US has allocated the licensed band of 5.9 GHz (5.85-5.925 GHz) to 802.11p standard.
    The cellular-base solution is developed based on LTE standard, and the 3GPP group has completed the initial cellular V2X standard in the Release 14. 
    Cellular-based VCNs can support both direct vehicle communications (such as V2V and V2P) and cellular communications with networks (mainly V2I).
    Specifically, the enhanced vehicle-to-everything (eV2X) use cases have been defined to support diverse vehicular communication services, including enhanced mobile broadband (eMBB), ultra-reliable and low-latency communication (URLLC), and massive machine-type communication (mMTC) \cite{3GPP_V2X}.
    
    IEEE 802.11p and cellular-based solutions exhibit different advantages, and the future VCNs
    are envisioned to be an integration of both techniques \cite{Khadige16_DSAC_cellular_integration_survey_TVT}. 		
    IEEE 802.11p standard has established the foundation of VCNs and is ready to deploy, which is well accepted in US with commercially available hardware like on-board units (OBUs) and RSUs. 
    Although lack of commercial chips and spectrum for now, the cellular-based LTE-V is still considered competitive. 
    Without integration barriers, LTE-V can enjoy the ubiquitous coverage of existing infrastructures, comprehensive state-of-the-art technical supports, and high penetration rate with large population of users.
    In addition, the LTE-V solutions have been shown to achieve better performance than the contention-based 802.11p standard in terms of capacity, reliability and latency, specially in dense urban scenarios \cite{Ericsson_LTEV}.
    As a key use case of the next generation (5G) wireless networks, LTE-V is now under live trail test stage in China and European countries.		
    
    \subsection{Driving Intelligence with VCNs}
    
    \begin{table}
    	\footnotesize
    	\newcommand{\tabincell}[2]{\begin{tabular}{@{}#1@{}}#2\end{tabular}}
    	\centering
    	\caption{VCN context infomration classification in automated driving}
    	\label{tab_context_V2X}		
    	\begin{tabular}{|c|c|c|c|c|c|}
    		\hline
    		\multicolumn{2}{|c|}{Catergory} & \multicolumn{2}{|c|}{Contents} & Requirements & V2X modes\\
    		\hline
    		\multirow{4}{*}{\tabincell{c}{Driving \\ related}} & \multirow{3}{*}{\tabincell{c}{Micro \\ scale}} & \multirow{1}{*}{\tabincell{c}{Surrounding \\ vehicles}} & \tabincell{c}{Position, mobility (speed, steering, headway), \\ condition (gas, engine, lights, wheel pressure), \\ special vehicle type, route} & \multirow{2}{*}{\tabincell{c}{Periodic, \\ minimal 1-10 Hz,\\ delay $\leq$ 100 ms}} & \tabincell{c}{V2V, V2I} \\
    		\cline{3-4}
    		\cline{6-6}
    		& & Road condition & Map, signs, signals, constructions, restrictions & & \tabincell{c}{V2I, V2RS}\\
    		\cline{3-6}
    		& & Emergency & \tabincell{c}{Overtaking, lane changing, merge, brake alarm, \\ accident warning, pedestrians} & \tabincell{c}{Event-driven,\\ delay $\leq$ 100 ms} & \tabincell{c}{V2V, V2P}\\
    		\cline{2-6}
    		& \tabincell{c}{Macro \\ scale} & \multicolumn{2}{|c|}{Traffic density, traffic jam, parking lot, ETC} & Delay $\leq$ 500 ms & V2I, V2C \\
    		\hline
    		\multicolumn{2}{|c|}{Travel experience} & \multicolumn{2}{|c|}{V2V social networks, Internet access, location-based services} & Real-time/elastic & \tabincell{c}{V2V, V2I}\\
    		\hline
    		\multicolumn{2}{|c|}{Vehicle-as-sensor} & \multicolumn{2}{|c|}{City mapping, air pollution,  weather, and facility monitoring} & Real-time/elastic & \tabincell{c}{V2I, V2C}\\			
    		\hline
    	\end{tabular}
    \end{table}
    
    VCNs can bring more intelligence to automated driving, and thus address both the technical and commercial obstacles.
    
    \textbf{1. Individual Driving Intelligence}
    
    As discussed in the last section, sensing-based automated driving intelligence can be constrained by ranges and environments.
    VCNs can help to overcome these obstacles by providing global driving information in spatial and temporal domains.		
    Through camera sharing, each vehicle can obtain bird view of the whole driving path, even when the sight is blocked by large-size vehicles ahead. 
    With such information, vehicles can make globally optimal control decisions and take actions in advance.
    Related applications can be best route planning, parking lot searching, and fast starting when traffic lights turn green.
    
    The performance of automated driving intelligence can be also constrained by the limited on-board hardware or software capabilities, which can be addressed if vehicles can utilize the network resources through VCN outsourcing.
    High definition map is critical in automated driving but of intensive data size (5 GB/km).
    Although vehicles cannot store the complete map, they can always download the needed piece from cloud via VCNs.
    Additionally, the driving-irrelevant computing tasks (like entertainment) can be offloaded to edge cloud, which can relieve the on-board processing resources and accelerate the speed of automated driving system.

    \textbf{2. Cooperative Swarm Intelligence}
    
    With VCN-enabled interactions, vehicles on the road can coordinate to enhance safety and efficiency, as well as form a swarm intelligent system.
    From the safety perspective, vehicles can share their driving intentions which cannot be observed through sensing techniques, such that neighboring vehicles can be better prepared in advance to overcome challenging driving tasks like lane changing, overtaking, merge maneuvers and emergent braking.
    In case of emergent braking, the vehicles can send messages to following cars immediately, to avoid rear-end collisions.	
    From the efficiency perspective, VCN-assisted cooperative vehicle platooning can greatly increase transportation efficiency in terms of fuel consumption, navigation, and road utilization with shorter headway.
    Furthermore, by replacing traffic lights by VCN-based controllers at intersections, fine-grained vehicle-level control can be optimized based on the instant traffic conditions, which holds the promise to increase transportation efficiency and reduce waiting time.
    
    Furthermore, vehicles can obtain in-depth understanding of the driving environment, by combining the information of neighboring devices.
    One example can be high accuracy positioning, which can be achieved by combining absolute positioning (i.e., GPS signals) and relative positioning (i.e., RSU signals, smart phones and other vehicles).  
    The received VCN information can also be utilized for error prediction, detection and correction, in case of software or hardware failures.
    
    \textbf{3. Friendly Serving Intelligence}
    
    In addition to improved intelligence quotient (IQ), VCNs also enhance the emotional quotient (EQ) of automated driving presented to customers, by providing enriched service experience, reducing transportation costs, as well as promoting smart city evolutions.
    
    The VCN-enabled Internet access in automated driving cars can bring revolutionary change to the transportation styles, with the emerging on-road mobile applications like augment reality, virtual reality, and location-based services.
    The customers can get on or off automated driving cars anywhere based on their need without parking or walking, but only through on-line registration.
    Furthermore, the automated driving cars can become a customized mobile home, office, game center or restaurant, in addition to a pure transportation tool.			
    
    The VCN-assisted transportation information acquisition can enable new business models like efficient automated car sharing, bringing potential benefits like higher vehicle utilization, reduced parking lots, and lower manufacture wastes.
    In addition, the sensor-rich vehicles can share the collected data to serve smart city construction via on-line or off-line V2I communications, such as high definition city mapping, air pollution and weather monitoring, and facility failure report.
    
    \subsection{Related Research Works}
    
    Currently, the research of automated driving mostly focuses on the sensing intelligence enhancement, whereas very few works have considered the combination of VCNs.
    In this part, we review several recent works on typical VCN-assisted automated driving applications.
    
    Automated driving vehicles should be capable to detect different lanes, in driving tasks of ACC, lane changing or merge maneuvers.
    Although computer vision techniques can be applied, it may fail under bad road condition with faded lanes.
    High accuracy vehicle positioning is considered as a supplementary or even alternative solution in this case, which requires centimeter-level accuracy.
    Conventional absolute positioning technologies, inertial navigation aided by a Global Navigation Satellite System (GNSS) (eg., US Global Positioning System (GPS)), are insufficient due to the meter-level accuracy and constrained availability in scenario of tunnels.
    To address these challenges, several works have proposed VCN-assisted relative positioning methods.
    Aly \emph{et al.} have designed the LaneQuest lane detection system by utilizing the inertial sensors of ubiquitous smartphones in a crowd-sourcing manner, which can achieve accurate lane detection 80\% of the time with high energy-efficiency \cite{lane_detection_phone_conf15}. 
    Wang \emph{et al.} have suggested to exploit the physical layer information of V2V communications (such as signal strength and range-rate) to conduct cooperative positioning, which is estimated to improve the positioning performance by 35\%-72\% depending on the vehicle density, compared with the standalone GPS method \cite{co_position_Access16}.	
    
    VCN-assisted cooperative adaptive cruise control (C-ACC) and vehicular platooning have also drawn extensive research extension.
    In conventional ACC, the information of vehicles in front are obtained through sensing technologies and utilized for feedback loop control of the following vehicles.
    V2V communications further provide the preceding vehicle's acceleration information, whereby C-ACC can conduct longitudinal automated vehicle control with a feed-forward loop.
    Related works have shown C-ACC is able to shorten inter-vehicle distance with safety and improve the stability of vehicle strings, and thus save fuel consumption with reduced velocity variance and aerodynamic drag \cite{CACC_ITIS06}.		
    Furthermore, the joint design of the V2V communication topology and vehicle following control can further enhance the stability of automated vehicle strings in frequency-spatial domain \cite{CACC_codesign_TITS}.
    Automated vehicle platooning goes beyond C-ACC, where a group of vehicles move at the same speed in proximity with a lead vehicle and a number of followers vehicles \cite{CACC_mag15}.
    As comparison, the required inter-vehicle intervals for human driving, conventional on-board ACC (without V2X communication), and platooning are 3 seconds, 1.6 seconds, and 0.5 second (corresponding to 100 m, 53 m, and 17 m at speed of 120km/h), respectively \cite{CACC_mag15}.			
    In addition to efficiency, C-ACC has been also applied to detect on-board sensor faults and function failures by jointly processing and analyzing the obtained data of neighboring vehicles \cite{CACC_detection_conf16}.
    These detected results can be applied to predict potential hardware and software failures, and thus enhance the reliability of automated driving system by taking measures in advance.

\section{Case Studies}
    \label{sec_CaseStudy}
\begin{figure}[t]
    	\centering
    	\includegraphics[width=0.40\linewidth]{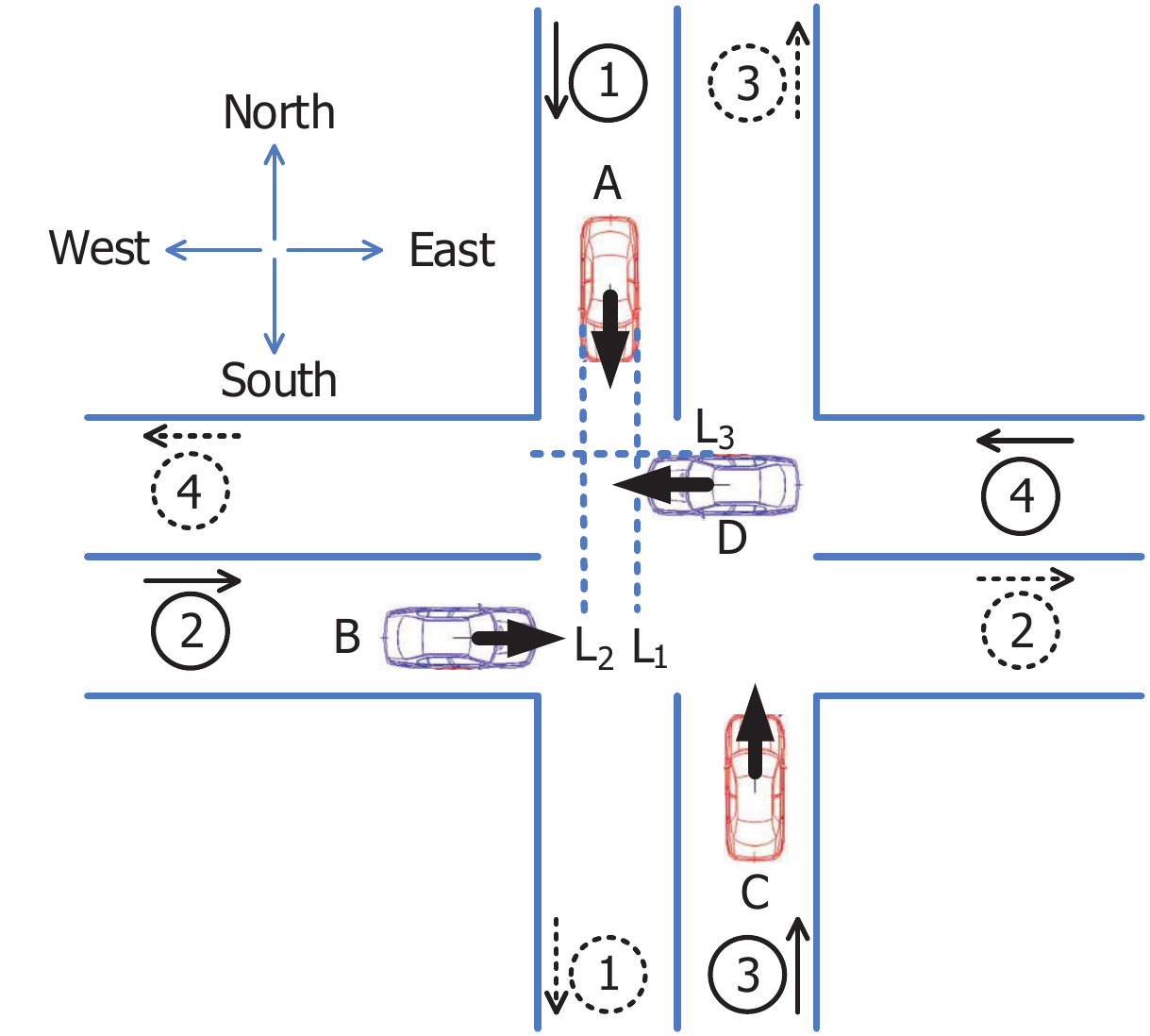}
    	\caption{{{V2V-based intersection passing illustration.}}}
    	\label{intersection}
    \end{figure}

    This section presents a case study where communication and sensing techniques are coordinated to support automated vehicles passing an intersection without the traffic light control.
    {{A four-direction intersection is considered and each direction has a road segment of 4 kilometers, as shown in Fig.~\ref{intersection}. 
    		The four directions are marked by \ncircled{1} \ncircled{2} \ncircled{3} and \ncircled{4}, respectively.
    		Vehicles arrive in each direction, go straight through the intersection and depart at the end of road segment.
    		In each direction, the inter-arrival time between two neighboring vehicles follows uniform distribution of $[T - 0.5 s, T + 0.5 s]$, where $T$ denotes the average inter-arrival time depending on vehicle traffic density.
    		Vehicles are set with the same performance parameters, i.e., length = 5 m, width = 2 m, maxSpeed = 15 m/s, accelerate = 10 m/s$^{2}$, decelerate = 10 m/s$^{2}$.
    		Suppose all vehicles move at the maximum speed when entering the road segments, but dynamically adjust the speed to maintain a minimal headway of 30 m considering safety and efficiency.}}
    
    Two methods are compared, i.e., conventional traffic light-based control and V2V-based control.
    {{For the light-based control, vehicles wait for the green light to pass the intersection, where the green, yellow, and red light durations are set as 30 s, 3 s and 33 s, respectively. 
    		For the V2V-based control, vehicles keep up-to-date knowledge of surrounding vehicles (including position, velocity, heading direction and acceleration), and decide whether to brake or pass when bounding to entering the intersection in a distributed manner.
    		Fig.~\ref{intersection} gives an illustration, supposing four vehicles arrives the interaction in order of $D$, $B$, $C$, $A$.
    		For example, vehicle $A$ needs to calculate if vehicle $D$ can completely pass the dashed line $L_{2}$ before $A$ arrives at the dashed line $L_{3}$.}}
    Simulations are conducted in Python, where each vehicle obtains the kinematic information and computes the decision results every 100 ms. 

    \begin{figure*}[t]
    	\centering
    	\subfloat[CDF of delay with even traffic]{
    		\includegraphics[width=0.4\linewidth]{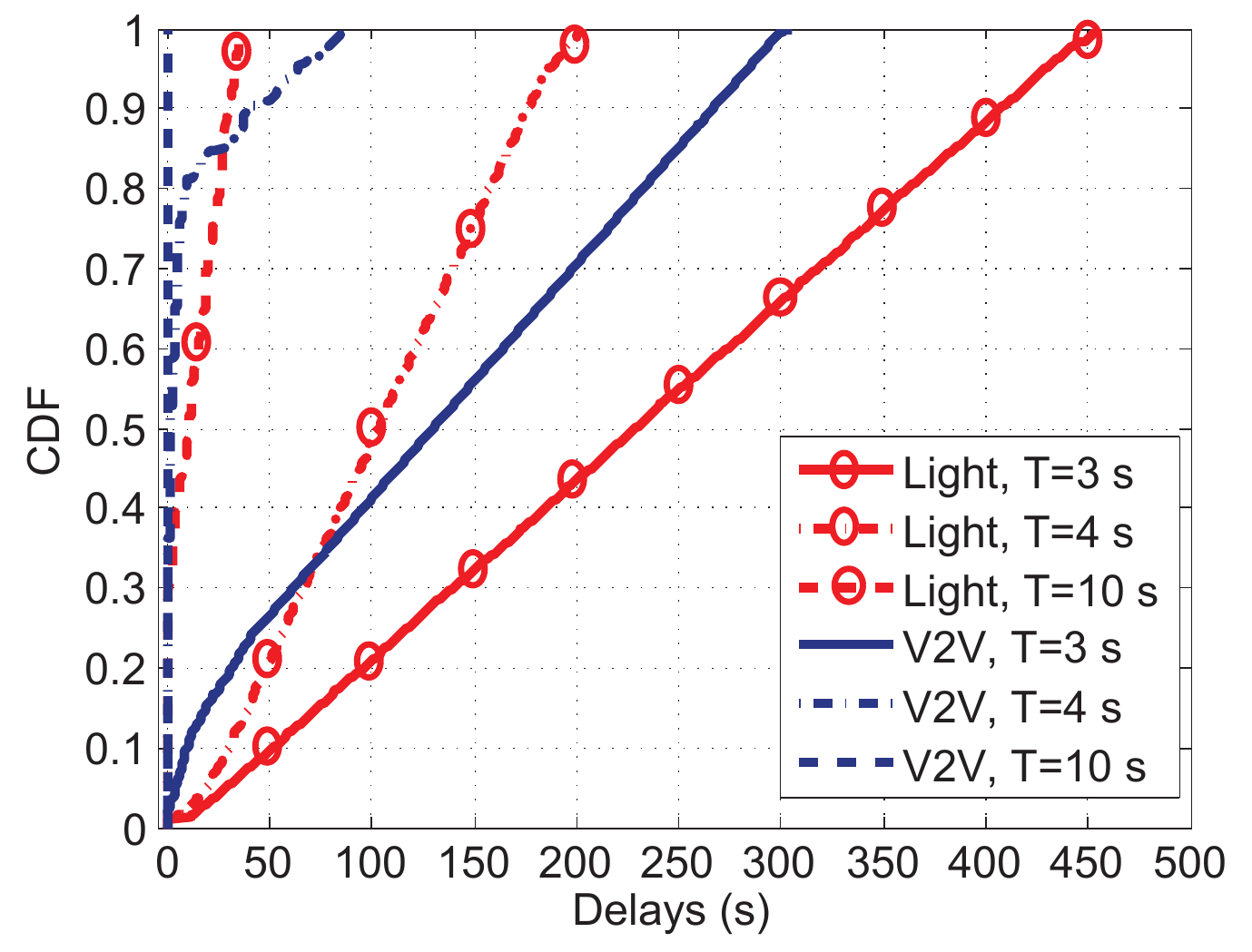}}
    	\subfloat[CDF of delay with uneven traffic]{
    		\includegraphics[width=0.4\linewidth]{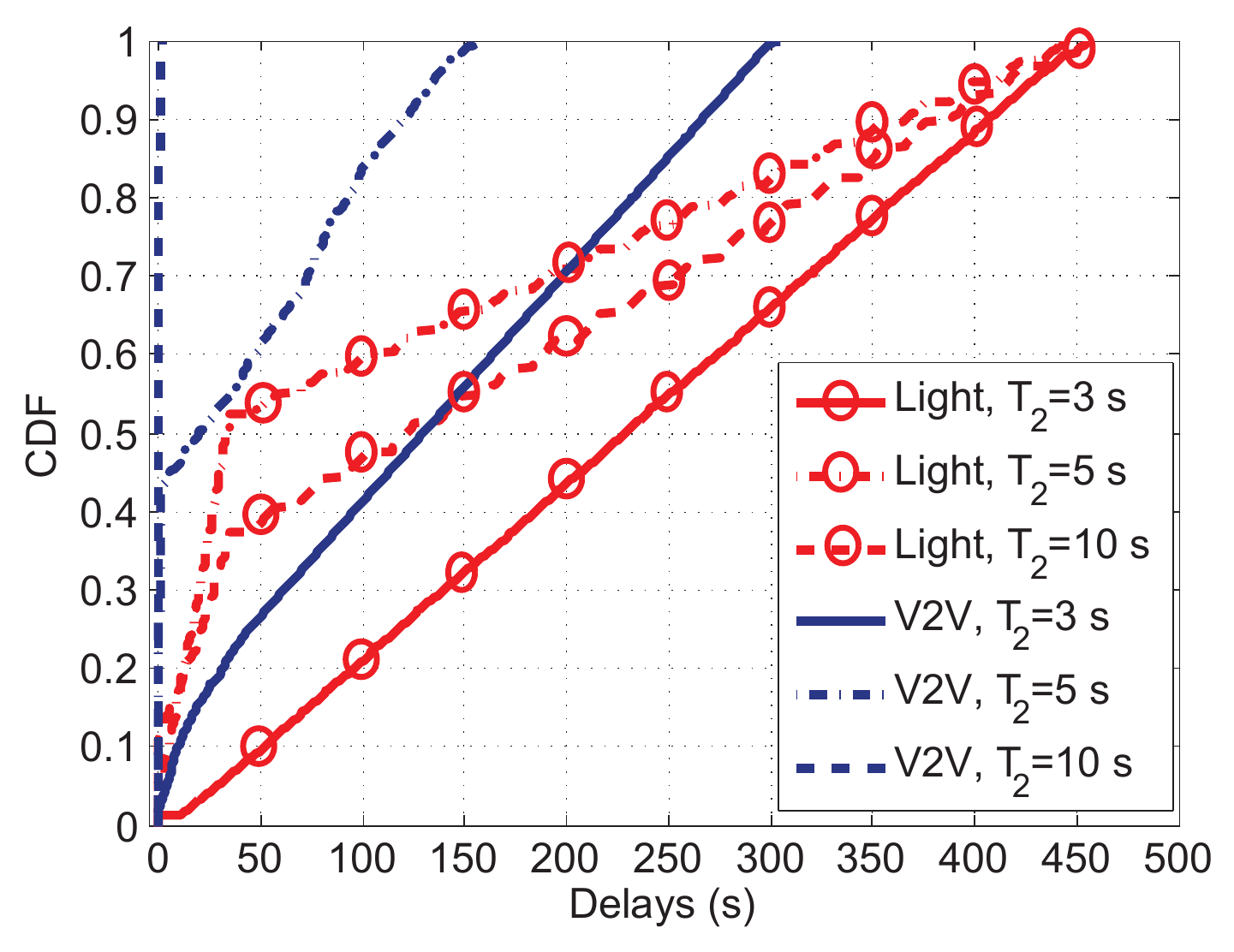}}
    	\caption{{{Intersection delay performance comparison}}.}
    	\label{fig_performance}
    	\vspace{-0.5cm}
    \end{figure*}

    {{The delay performances of light- and V2V-based schemes are illustrated in Fig.~\ref{fig_performance}~(a), when the vehicle arrival rates are equal in both the East-West and South-North directions.
    		Three facts can be observed from Fig.~\ref{fig_performance}~(a). }}
    First, the V2V-based scheme always surpasses the light-based scheme regardless of traffic load.
    {{For illustration, the median delay (i.e., with the CDF values of 0.5) of the light-based scheme is 230, 100 and 20 s when the vehicle inter-arrival time $T$ is $3$, $4$ and $10$ s, respectively.
    		These figures will reduce to 130, 5 and 0 s by adopting the V2V-based scheme.}}
    Second, the V2V-based scheme can eliminate the intersection delay at low vehicle arrival rate, whereas vehicles always suffer from intersection delay under the light-based scheme.
    For instance, when $T=10$ s, $30\%$ vehicles experience a delay longer than about $20$ s under the light-based scheme, whereas all vehicles pass the intersection without delay under the V2V-based scheme.
    Third, the light-based scheme is more sensitive to the traffic density especially at low vehicle arrival rates.
    For example, when $T$=4 s, 30\% vehicles experience a delay longer than 140 s, while this value will surprisingly increase to 320 s when $T$=3 s.
    The reason is that the following vehicles need to guarantee the minimal headway, and thus the intersection delay will accumulate rapidly as the traffic load increases.
    Since the V2V-based scheme introduces cooperations, it can significantly improve the intersection efficiency to reduce delay, which is especially favorable when the traffic load is heavy.
    
    {{Fig.~\ref{fig_performance}~(b) illustrates a more general case with uneven vehicle arrival rates in different directions.
    		Specifically, the vehicle inter-arrival time in the South-North direction $T_1$ is fixed to 3 s, while the vehicle inter arrival time in the East-West direction $T_2$ varies to depict the degree of traffic unevenness.
    		The results show piece-wise shapes due to the traffic unevenness, where the first piece corresponds to the under-utilized East-West direction and the second piece corresponds to the congested South-North direction.
    		The simulation results show that the V2V-based scheme can significantly reduce the intersection delay under all traffic load settings.
    		In addition, as the traffic density at East-West direction decreases, the traffic congestion problem at South-North direction can be mitigated under the V2V-based scheme, whereas the problem remains under the light-based scheme.
    		The reason is that the V2V-based scheme can conduct fine-grained vehicle scheduling based on the real-time vehicle mobility information, improving intersection utilization and efficiency.
    		On the contrary, the light-based scheme is not flexible to traffic variations, and thus the intersection can be under-utilized due to traffic unevenness, introducing high intersection delay.
    		This preliminary experiment reveals the attractive potential of V2V communications on improving road efficiency.
    		Future work can design more efficient vehicle scheduling algorithms considering different practical transportation scenarios, such as turning vehicles, V2X communication failures, and machinery control inaccuracy.}}
    
\section{Open Research Issues}
    \label{sec_research_topic}

    The sensor-rich automated vehicles (100-300 sensors per car) are expected to generate huge amount of data (GigaBytes per second), considering the frequent V2V interaction (minimum 1-10 times per second) and data-hungry applications (HD map, variety location-based services, and augment reality).
    Additionally, the safety-related applications have posed requirements of extremely high reliability and low latency.
    {{However, the mobility of vehicles brings highly dynamic network topologies and fast varying channel conditions.
    		Therefore, VCNs face great challenges when dealing with the contradiction between strict QoS demands and unreliable V2X resources.
    		In this section, we highlight interesting research topics.}}
    
    \subsection{Integrated Vehicular Networking}
    
    {{Existing work has proposed to make use of the unlicensed WiFi or IEEE 802.11p spectrum for opportunistic vehicular mobile traffic offloading, considering the heavy load of cellular networks \cite{Cheng16_WIFI_offloading_TITS}.
    		In the future, more networks with complementary advantages can join this cooperation and form the advanced space-air-ground integrated networking \cite{NZhang_SAG_mag17}.
    		Specifically, ground networks (including V2V, cellular base stations and RSUs) can provide high capacity to support massive connections through extensive spectrum reuse.
    		Air facilities (drones and high altitude platforms) can provide favorable line-of-sight (LOS) connections as relays in large coverage, and move with vehicles to maintain handover-free access.
    		Satellites can provide positioning and navigation services, and compensate ground network coverage in areas like countrysides and mountains.
    		However, different networks currently use different technologies, introducing challenges to integration.
    		Software-defined network is a potential solution, which can coordinate different networks in the control plane through a logically centralized controller.
    		The design of control plane is still under-developed, and how to implement the control decisions in the data plane is also an important issue.}}
    
    {{	
    		\subsection{Co-Design of Computation and V2X Communications}
    		Future VCNs are expected to provide computation as a service to assist automated vehicles making fast and accurate decisions. 
    		Specifically, mobile edge computing provides vehicles with rich computation resources in proximity, whereby the computation-overloaded vehicles can offload tasks to neighboring vehicles or RSUs for driving performance enhancement \cite{Chen17_D2D_MEC}.
    		As the performance relies on both computation and communication efficiency, mobile edge computing calls for the co-design of computation and V2X communication scheduling.
    		This problem can be extremely challenging considering the integrated network heterogeneity, V2X channel unreliability, vehicle mobility and dynamic computation resource occupancy \cite{Chang17_delay_multi_VANET}.
    		To deal with these issues, vehicles can apply advanced machine learning method to obtain the channel information and resource availability, and then choose the best neighbor to offload computation tasks \cite{Sun17_Ve_offloading_JSAC}.
    		Meanwhile, critical computation tasks can be replicated and offloaded to multiple neighbors to guarantee reliability. 
    		In addition, incentive schemes should be designed to encourage more vehicles to provide assistance.}}
    
    \subsection{Big Data in Vehicular Networking}
    
    Although the big data generated in automated driving bring burdens for communication, computing and storage, they also provide opportunities to improve VCN optimization.
    Specifically, big data analysis provides in-depth understanding of on-road mobile traffic demand and network status.
    {{For example, the large-scale vehicle mobility knowledge (e.g., spatial vehicle density) can be utilized to guide vehicular network infrastructure planning, while the small-scale vehicle mobility information (e.g., headway and speed) can be exploited in vehicle communication scheduling.
    		The sensor-rich automated driving vehicles can cooperate to analyze the instant network status.}}
    In this regard, existing work has applied location-based crowd-sourcing to predict the nearest RSU location, based on which efficient vehicle communication routing schemes are designed to enhance vehicular network reliability \cite{crowdsourcing_Vehi_routing_TITS13}.
    {{Similarly, many design issues can be revisited to discover new opportunities for intelligent network management. }}

    \subsection{Security and Privacy}
    Vehicular networking can bring significant benefit to automated driving intelligence, but may also become a vulnerable part and pose security and privacy issues.
    The sensor-rich and V2X-enabled automated driving vehicles can easily leak out personal information, {{and even become }}a dangerous bomb on-road once hacked.
    Possible solutions include data encryption, physical layer security analysis, security and privacy protocol design. 
    However, security and privacy provisioning may degrade the network efficiency.
    For example, authentication, a key process to guarantee network security, may bring significant delay during V2X communications.
    {{Accordingly, how to secure the V2X communications with transmission efficiency is also critical, especially for the delay-sensitive V2X applications.}}
  
    
\section{Conclusions}
	\label{sec_conclusions}

	This article has briefly reviewed the state-of-the-art sensing-based automated driving and VCN technologies, and probed into the necessities and potentials to conduct joint research.
	Specifically, we have discussed how VCNs can enhance the performance of sensing-based automated driving vehicles, by providing global information, enabling vehicle interaction and cooperation, as well as developing attractive cloud-based mobile services and applications.
	A case study on intersection passing has demonstrated the potential of VCN-assisted automated driving on improving transportation efficiency.
	Furthermore, several critical research issues have been highlighted for future in-depth studies from the VCN perspective.
	Nevertheless, the integration of networking and automated driving raises potential interdiscipline research of varied subjects in addition to communication and information technologies, such as on-road application and service development, business models, and transportation system rebuilding.
	Joint research efforts can be devoted to accelerate the launch of automated driving era.


\end{document}